\pdfoutput=1

\documentclass[11pt]{article}

\usepackage[final]{acl}

\usepackage{times}
\usepackage{latexsym}

\usepackage[T1]{fontenc}

\usepackage[utf8]{inputenc}

\usepackage{microtype}

\usepackage{inconsolata}

\usepackage{graphicx}

\usepackage{comment}
\usepackage{xspace}
\usepackage{graphicx}
\usepackage{caption}
\usepackage{multirow}
\usepackage{multicol}
\usepackage{makecell}
\usepackage{afterpage}
\usepackage{booktabs}
\usepackage{tabularx}
\usepackage{adjustbox}
\usepackage{pifont}
\usepackage{soul}
\usepackage{enumitem}
\usepackage{amsmath}
\usepackage{amssymb}
\usepackage{comment}
\usepackage{enumitem}
\usepackage{cleveref}
\usepackage{arydshln}
\usepackage{graphicx,dblfloatfix}
\usepackage{algorithm}
\usepackage{algpseudocode}
\usepackage{amsmath}
\usepackage{dsfont}

\title{Can You Share Your Story? Modeling Clients' Metacognition and Openness for LLM Therapist Evaluation}

\author{
        Minju Kim\textsuperscript{\rm 1}
        Dongje Yoo\textsuperscript{\rm 1}
        Yeonjun Hwang\textsuperscript{\rm 1}\\
        \textbf{Minseok Kang\textsuperscript{\rm 1}}
        \textbf{Namyoung Kim\textsuperscript{\rm 1}}
        \textbf{Minju Gwak\textsuperscript{\rm 1}}\\
        \textbf{Beong-woo Kwak\textsuperscript{\rm 1}}
        \textbf{Hyungjoo Chae\textsuperscript{\rm 1} }
        \textbf{Harim Kim\textsuperscript{\rm 1}}
        \textbf{Yunjoong Lee\textsuperscript{\rm 2}}\\
        \textbf{Min Hee Kim\textsuperscript{\rm 1}}
        \textbf{Dayi Jung\textsuperscript{\rm 1}}
        \textbf{Kyong-Mee Chung\textsuperscript{\rm 1}}
        \textbf{Jinyoung Yeo\textsuperscript{\rm 1}\thanks{Corresponding author}} \\
        Yonsei University$^{1}$ Eulji University$^{2}$\\
        \texttt{\{minnju,jinyeo\}@yonsei.ac.kr}
}

\begin{document}
\maketitle
\newcommand{\todoc}[2]{{\textcolor{#1}{#2}}}
\newcommand{\todoblue}[1]{\todoc{blue}{#1}}

\newcommand{\todocc}[2]{{\textcolor{#1}{[[#2]]}}}
\newcommand{\todored}[1]{\todocc{red}{[[#1]]}}
\newcommand{\hist}[1]{\todored{hist: #1}}

\newcommand{\todocccc}[2]{{\textcolor{#1}{[[#2]]}}}
\newcommand{\todogreen}[1]{\todocccc{green}{[[#1]]}}
\newcommand{\minju}[1]{\textcolor{brown}{#1}}
\newcommand{\red}[1]{\textcolor{red}{#1}}

\newcommand{\se}{{\it SE}}%
\newcommand{\eg}{{\it e.g.}}%
\newcommand{\ie}{{\it i.e.}}%
\newcommand{\etal}{{\it et al.}}%
\newcommand{\etc}{{\it etc}}%
\newcommand{\ourframework}{\textsc{Clam}\xspace}
\newcommand{\ourdata}{\textsc{Pearl}\xspace}

\newcommand{\argmin}{\operatornamewithlimits{argmin}}
\newcommand{\argmax}{\operatornamewithlimits{argmax}}
\newcommand{\tabitem}{~~\llap{\textbullet}~~}
\definecolor{yellow-green}{rgb}{0.3, 0.5, 0.0}

\definecolor{lightergray}{RGB}{230,230,230}
\definecolor{DarkGreen}{RGB}{30,130,30}
\newcommand{\cmark}{\textcolor{DarkGreen}{\ding{51}}}
\newcommand{\xmark}{\textcolor{red}{\ding{55}}}%


\newcommand{\mcal}[1]{{\cal{#1}}}
\newcommand{\calA}{\mbox{${\cal A}$}}
\newcommand{\calB}{\mbox{${\cal B}$}}
\newcommand{\calC}{\mbox{${\cal C}$}}
\newcommand{\calD}{\mbox{${\cal D}$}}
\newcommand{\calE}{\mbox{${\cal E}$}}
\newcommand{\calF}{\mbox{${\cal F}$}}
\newcommand{\calG}{\mbox{${\cal G}$}}
\newcommand{\calH}{\mbox{${\cal H}$}}
\newcommand{\calI}{\mbox{${\cal I}$}}
\newcommand{\calJ}{\mbox{${\cal J}$}}
\newcommand{\calK}{\mbox{${\cal K}$}}
\newcommand{\calL}{\mbox{${\cal L}$}}
\newcommand{\calM}{\mbox{${\cal M}$}}
\newcommand{\calN}{\mbox{${\cal N}$}}
\newcommand{\calO}{\mbox{${\cal O}$}}
\newcommand{\calP}{\mbox{${\cal P}$}}
\newcommand{\calQ}{\mbox{${\cal Q}$}}
\newcommand{\calR}{\mbox{${\cal R}$}}
\newcommand{\calS}{\mbox{${\cal S}$}}
\newcommand{\calT}{\mbox{${\cal T}$}}
\newcommand{\calU}{\mbox{${\cal U}$}}
\newcommand{\calV}{\mbox{${\cal V}$}}
\newcommand{\calW}{\mbox{${\cal W}$}}
\newcommand{\calX}{\mbox{${\cal X}$}}
\newcommand{\calY}{\mbox{${\cal Y}$}}
\newcommand{\calZ}{\mbox{${\cal Z}$}}

\definecolor{lightblue}{RGB}{224,236,247}
\definecolor{deepblue}{RGB}{9,46,107}

\newcommand{\subscript}[2]{$#1 _ #2$}

\crefformat{section}{\S#2#1#3}
\crefformat{subsection}{\S#2#1#3}
\crefformat{subsubsection}{\S#2#1#3}
\crefrangeformat{section}{\S\S#3#1#4 to~#5#2#6}
\crefmultiformat{section}{\S\S#2#1#3}{ and~#2#1#3}{, #2#1#3}{ and~#2#1#3}

\newcommand{\ours}{\textsc{MindVoyager}\xspace}

\begin{abstract}
\red{This work does NOT advocate for the use of large language models (LLMs) in psychological counseling. Instead, we propose an assessment approach to reveal the characteristics of LLM therapists.}
Understanding clients’ thoughts and beliefs is fundamental in counseling, yet current evaluations of LLM therapists often fail to assess this ability. Existing evaluation methods rely on client simulators that clearly disclose internal states to the therapist, making it difficult to determine whether an LLM therapist can uncover unexpressed perspectives. To address this limitation, we introduce \ours{}, a novel evaluation framework featuring a controllable and realistic client simulator which dynamically adapts itself based on the ongoing counseling session, offering a more realistic and challenging evaluation environment. We further introduce evaluation metrics that assess the exploration ability of LLM therapists by measuring their thorough understanding of client's beliefs and thoughts.

\end{abstract}

\section{Introduction}

Recent advances in large language models (LLMs)~\citep{openai2023gpt, dubey2024llama} have sparked interest in their potential applications for mental health and well-being~\citep{nie2022conversational,hauser2022promise}. One promising area is the development of LLM therapists, which are dialogue agents designed to conduct counseling sessions with clients interactively and provide therapeutic support and guidance~\citep{lee2024cactus,xiao-etal-2024-healme,na-2024-cbt}. 

\begin{figure}[t!]
    \centering
    \includegraphics[width=\linewidth]{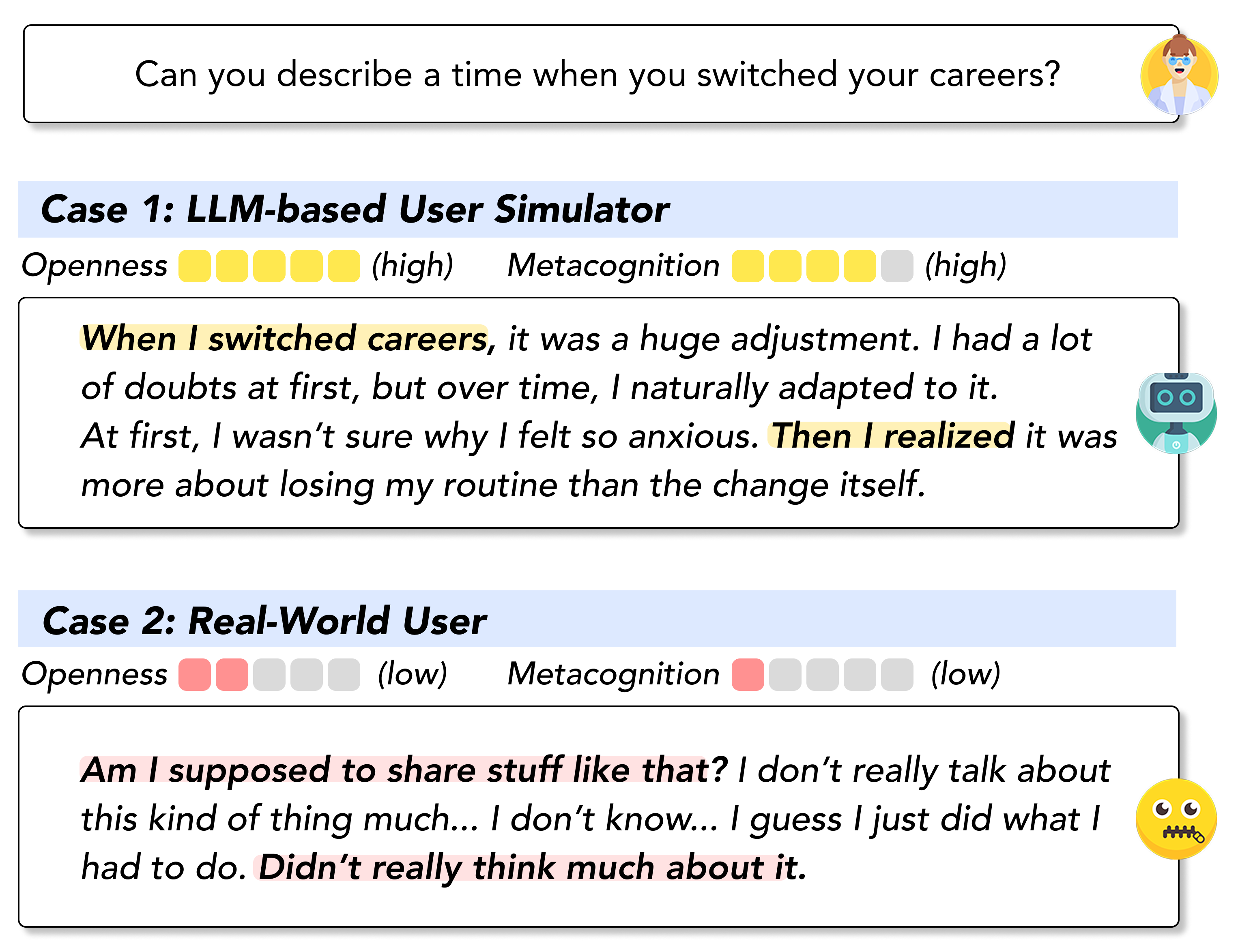}
    \caption{LLM-based user simulator vs. Real-world user: User simulators are more prone to act overly compliant and self-aware in therapeutic sessions.}
    \label{fig:intro_figure}
\end{figure}
While there are various psychological counseling paradigms, the most fundamental and foundational stage in all paradigms is the \textbf{exploration stage}, where the therapist and client collaboratively uncover the client's experiences, emotions, and thought patterns~\citep{corey1991theory,meier1924elements,carey2004socratic}. The exploration stage enables the therapists have a deep understanding of the client and provide proper guidance ~\citep{wang2024towards,qiu2024interactive}. However, existing evaluation methods for psychological counseling primarily assess whether LLM therapists can help the client feel better and provide effective solutions, neglecting the importance of the exploration ability of LLM therapists~\citep{wang2024towards,xiao-etal-2024-healme,lee2024cactus,na2024multi,na-2024-cbt, ji2023rethinking, lee2024cactus}. This gap arises due to the reliance on naive client simulators utilized for evaluation, which possess complete openness to the therapist and awareness of their situation, leading to articulating their experiences, emotions, and thought patterns early in the session, thereby eliminating the need for exploration.

To address these evaluation challenges, a realistic client simulator is needed which can emulate realistic client behaviors. Specifically, there are two notable differences between real-world clients and simulated clients: (1) \textbf{Openness} – willingness to share personal information and consider alternative perspectives. Clients with high openness often disclose experiences and actively engage the exploration stage, while those with low openness tend to be guarded and reluctant to share. (2) \textbf{Metacognition} – awareness of their own thoughts, emotions, and cognitive processes, as well as their ability to recognize throughout the exploration stage. A client with high metacognitive abilities can articulate themselves, whereas a client with low metacognition may struggle to recognize or verbalize their emotional states, leading to more stilted or guarded interactions. These two factors determine the difficulty of the experimentation stage, thus, client simulators aiming for real-world scenarios must incorporate these traits to effectively evaluate LLM therapists.

However, LLMs are typically optimized for fluent, coherent text generation, which can produce overly cooperative responses by default. As a result, simulating a realistically resistant or unaware client (\ie, client with low openness and low metacognition) is non-trivial. LLMs often fill in gaps with more detail than a reluctant, self-unaware client might provide, or they may inadvertently shift to cooperative, high-disclosure styles once the conversation becomes complex. Hence, prompt engineering alone is insufficient for simulation of challenging clients.

Motivated by these limitations, we propose a novel evaluation framework that utilizes a \emph{controllable} and \emph{realistic} client simulator for assessing the exploration ability of LLM therapists. Our framework features a client simulator with adjustable levels of \textit{openness} and \textit{metacognition}. To closely emulate authentic client behaviors, our client simulator dynamically adjusts openness throughout the session in response to the rapport established with the LLM therapist.  Also, depending on the level of metacognition, the speed at which the client simulator can identify their negative thoughts and core beliefs varies. With our framework, we evaluate whether LLM therapists can help clients recognize and articulate their own experiences and thoughts. Specifically, LLM therapists conduct counseling sessions with our client simulator and are evaluated by comparing their understanding of clients -- specifically core beliefs, intermediate beliefs, and potential early childhood memories -- and the ground-truth information. By focusing on the LLM’s capacity to explore thoughts and experiences, we aim to provide a robust measure of LLM therapists' potential as a therapeutic tool.

Our contributions are two-fold: (1) We are the first to introduce a client simulator that dynamically adjusts the openness and metacognition throughout the session based on the quality of the counseling session, offering a more realistic and challenging evaluation environment. (2) We propose a novel evaluation framework \ours{} that assesses the exploration ability of LLM therapists by comprehensively comparing their understanding on multiple aspects of the client, ensuring the necessity of the dynamic user simulator.


\section{Preliminaries}
Most existing methods for evaluating LLM therapists have adopted a self-play based approach, which utilizes LLMs as client simulators and simulates psychological counseling sessions by instructing LLM therapists and client simulators~\citep{wang2024towards,lee2024cactus}. While the client simulator plays a vital role in the evaluation methods, the simulation ability of LLMs has been overlooked. This raises concerns about self-play based evaluation methods and also LLM therapists evaluated by using naive client simulators. 
To this end, in this section, we analyze critical differences between LLM-based client simulators and real-world clients.

\subsection{Discrepancy between LLM-based Client Simulator and Real-world Clients}
To compare simulated client behavior to real-world clients, we conduct an analysis with three experts in psychological counseling. We collect twenty counseling sessions conducted between a client simulator and an LLM therapist. Here, we utilize GPT-4o-mini for both models. The experts are then asked to evaluate the sessions and provide qualitative feedback on discrepancies between simulated and real-world client responses. We systematically collect and analyze this feedback, identifying key areas of divergence.

\begin{figure}[t!]
    \centering
    \includegraphics[width=1.0\linewidth]{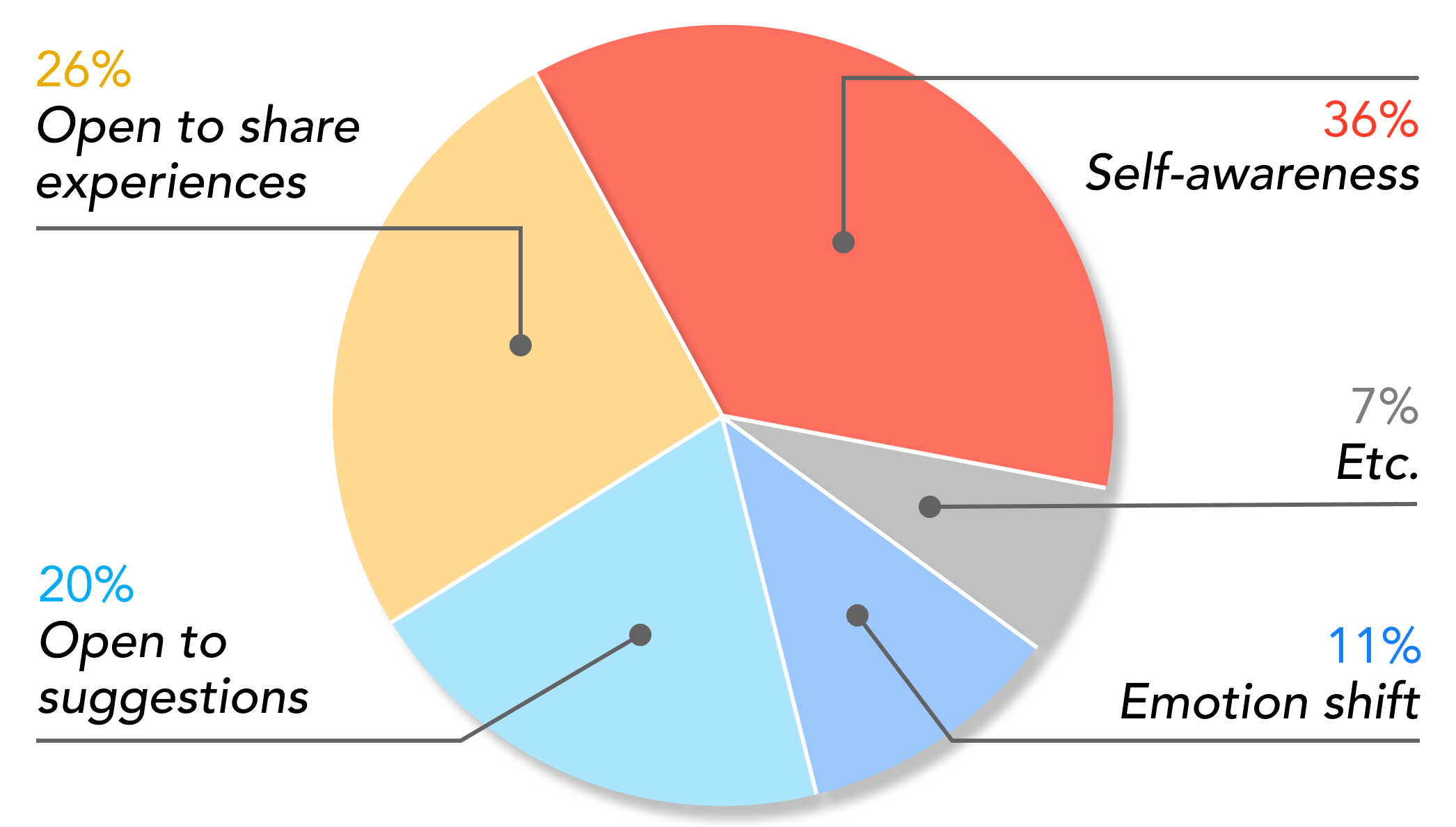}
    \caption{Key features of LLM-based client simulator differing from real-world clients, based on their responses in therapeutic sessions.}
    \label{fig:preliminary_chart}
\end{figure}

Figure~\ref{fig:preliminary_chart} summarizes feedback on the discrepancy between the client simulator and real-world clients. Overall, the feedback type percentages are 36.11\% for self-awareness, 26.39\% for openness to share experiences, 19.44\% for openness to any suggestion, and 11.11\% for rapid emotional transition. Experts summarize that, unlike real-world clients, the client simulator recognizes and explains feelings and behavior patterns clearly. In contrast, real-world clients do not know what to answer because they are confused, have not thought about it, or do not know where to start. Moreover, even if they know about themselves clearly, they have difficulty reporting in detail.

\subsection{Limitations of LLM-based Client Simulators}
As investigated in a previous analysis, metacognition and openness are the main differences between LLM-based client simulators and real-world clients. Here, we conduct additional analysis to find out whether prompt engineering can regulate the openness and metacognition of LLM-based client simulators. To this end, three prompts with different styles are collected and prompted to the GPT-4o-mini model to simulate clients. Then we compare the range of metacognition and openness of client simulators scaling from 1 to 5 by using a GPT-4o-mini model as a judge. We provide the prompts used for evaluation in  Appendix.

In our experiment, LLM-based client simulators instructed to behave as realistic clients exhibited high levels of openness and metacognition, achieving average scores of 5.0 in both traits. Even when prompted to display low openness and low metacognition, these simulators still demonstrated relatively high levels, with average scores of 4.28 and 4.15, respectively.

\section{\ours}
We propose \ours, a framework for assessing the exploration ability of LLM therapists utilizing a client simulator with adjustable metacognition and openness.



\subsection{Components of \ours}
Our client simulator consists of two main components: (1) the cognitive diagram, a structured framework that describes how an individual's thoughts and beliefs are interconnected and influence emotions and behaviors; and (2) the cognition mediator, which controls the accessible parts of the cognitive diagram throughout the counseling session.

First, we mask the cognitive diagram based on the client's metacognition and openness, then provide the masked cognitive diagram as input to an LLM to simulate a client. Throughout the counseling session, the cognition mediator evaluates the LLM therapist and progressively unmasks parts of the cognitive diagram, simulating a client gradually sharing thoughts and feelings.

\paragraph{Cognitive diagram.}
To develop a client simulator with realistic cognitive factors, such as beliefs and thoughts, we introduce a cognitive diagram inspired by the Cognitive Conceptualization Diagram (CCD). CCD is a standard framework in cognitive behavioral therapy (CBT) used to represent a client's cognitive structure~\citep{beck2020cognitive}.

To control the cognitive diagram, we categorize the elements from CCD into two types: external cognitive elements and internal cognitive elements, based on their depth within the human mind. For example, external cognitive elements include situations, thoughts, emotions, and behaviors, which can be easily recognized. In contrast, internal cognitive elements are typically harder to recognize and require deeper reflection to uncover—such as core beliefs, intermediate beliefs, coping strategies, and the relevant history that shaped those beliefs and strategies.

All elements are derived from CCD, and we provide definitions and examples in Table~\ref{tab:cognitive_model_explained}. Annotations for each element come from the Patient-$\psi$-CM dataset~\citep{wang2024patient}, which includes CCDs written by experts.

Formally, a cognitive diagram $G=E\cup I$ is defined as a union of two diagrams, which are external cognitive diagram $E$ and internal cognitive diagram $I$. External cognitive diagram $E=\{x_i,s_i\}$ includes external cognitive elements $x$ and associated status $s\in\{\texttt{masked, unmasked}\}$. Similarly, internal cognitive diagram $I=\{y_i,s_i\}$ consists of internal cognitive elements $y_i$ and masking state $s$.

\begin{figure*}[t!]
    \centering
    \includegraphics[width=\linewidth]{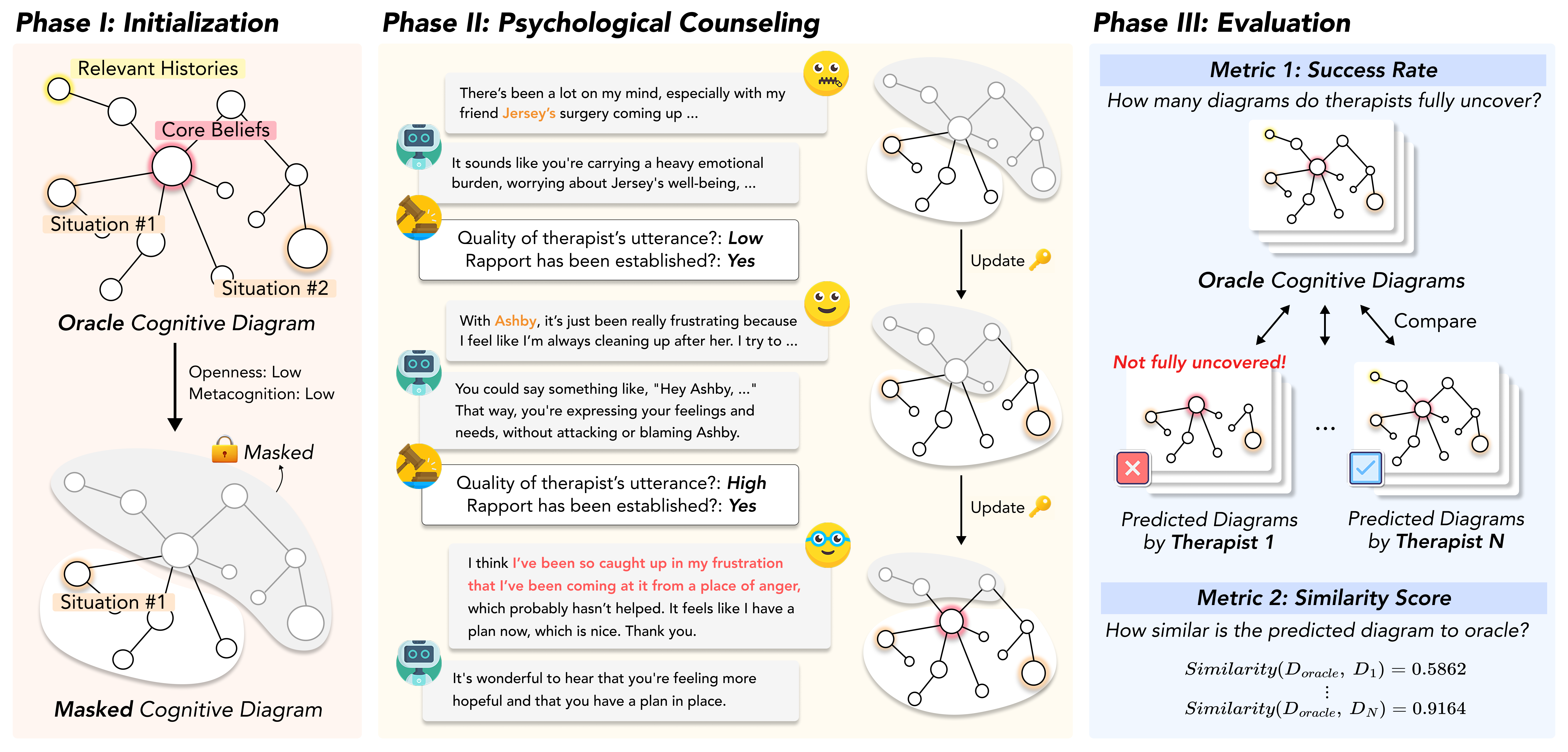}
    \caption{Overall procedure of \ours{}.}
    \label{fig:cognitive_model_construction}
\end{figure*}

\paragraph{Cognition mediator.}
While cognitive diagrams can provide realistic descriptions of thoughts, beliefs, emotions, and behaviors and explain how they influence each other, solely providing a cognitive diagram as input to an LLM to simulate a client can result in suboptimal simulations. LLMs typically generate cooperative responses that explicitly include most elements of the cognitive diagram, even when actual clients would not necessarily recognize or express them~\citep{zhou2024real}. To this end, we propose a cognition mediator that judges LLM therapists and manages the accessible part from the cognitive diagram based on the judgment. We provide detailed steps in the following section.

\subsection{Evaluation Process of \ours}
\paragraph{Phase I: Initialize cognitive diagram.} 
To simulate a client that gradually shares thoughts and beliefs, we first assign metacognition level and openness level to the simulator and mask part of the cognitive diagrams that should be unveiled throughout the session.

First, to make the client simulator unaware of elements in the internal cognitive diagram $I$, we mask all elements in the internal cognitive diagram $I' = \{(x_i,\texttt{masked})\}$ where $(x_i, s_i)\in I$. Also, based on the openness $o$ of the client, we manage the number of accessible elements $N$ in the external cognitive diagram $E' = \{(e_i,\texttt{unmasked})\}_{i<N}$, where $(e_i, s_i) \in E$.
After masking all elements in the internal cognitive diagram $I'$ and part of the external cognitive diagram $E'$, the masked cognitive graph $G'=\{E',I'\}$ is provided to the simulator.

\paragraph{Phase II: Simulate counseling session with LLM therapist.} 
Professional therapists can help clients discover their own beliefs and coping strategies, even when clients have low openness and low metacognition. To evaluate LLM therapists in realistic scenarios, the cognitive diagram of our client simulator is updated throughout the counseling sessions based on two main criteria: (1) Has rapport been established between the client and the LLM therapist? (2) Has the LLM therapist helped the client explore themselves?

Specifically, our cognitive mediator $M$ evaluates the ongoing session every $k$ turns to judge if the rapport has been established. Based on the ongoing session $sess$ and provided instruction $\text{Inst}_{rapport}$, cognitive mediator $M$ judges if the rapport has been established and determines the number of accessible elements in the external diagram $E$.
\begin{flalign}
    & rapport = Judge(sess, \text{Inst}_{rapport}) \\
    & N \gets N+1\quad \text{if } rapport == success \\
    & E' = \{(x_i,\texttt{unmasked})\}_{i<N}, \quad \forall(x_i,s_i) \in E
\end{flalign}
Moreover, based on the client's metacognition level, we set the number of turns, $l$, which determines how frequently the cognitive mediator. In every $l$ turn, the cognitive mediator $M$ evaluates the utterances from the LLM therapist given the ongoing session $sess$ and instruction $\text{Inst}_{exp}$. 
\begin{flalign}
    & exp = Judge(sess, \text{Inst}_{exp}) \\
    & I' = \{(y_i,\texttt{unmasked})\}_{i<|I|} \text{ if } exp == success 
\end{flalign}

\paragraph{Phase III: Evaluating LLM Therapists.} 
To assess the effectiveness of LLM-based therapists, we evaluate their performance after conducting counseling sessions with our client simulator. The evaluation is based on two key metrics: (1) Cognitive Diagram Exposure Rate (CDER) and (2) Induced Diagram Similarity Score (IDSS).

Cognitive diagram exposure rate (CDER) measures how well LLM therapists help clients recognize elements of the cognitive diagram. CDRS denotes the percentage of sessions the LLM therapist successfully reveals the entire cognition diagram during a counseling session. In this context, the metric is binary for each session—if the therapist reveals every element of the initially masked cognition diagram, the session is considered a success; if even one element remains concealed, it is considered a failure. The overall score is then calculated as the percentage of sessions in which the complete cognition diagram was successfully disclosed. 

Induced diagram similarity score measures the effectiveness of the LLM therapist in inducing the client simulator to elaborate on the revealed cognition diagram. Given the simulated counseling session, we first extract each element of the cognitive diagram by utilizing an LLM. Each element in the extracted cognitive graph is then compared with the corresponding element in the ground truth cognition diagram using semantic similarity scoring. This comparison quantifies how closely the content elicited in the session aligns with the intended cognitive structure.

Formally, we define these metrics as follows:
\begin{flalign}
    & \text{CDER} = \frac{\text{\# of sessions } G \text{ is revealed}}{\text{Total number of sessions}} \\
    & \text{IDSS} = semantic\_matching(I, I_{pred})
\end{flalign}
where $I$ represents the ground-truth internal cognitive diagram, and $I_{pred}$ is the internal cognitive diagram inferred by the LLM therapist. The function $semantic\_matching(\cdot, \cdot)$ measures whether each element of the diagram has the same semantic meaning, reflecting the degree to which the client elucidates elements in the cognitive diagram. Here, we do not consider the external cognitive diagram as the goal of psychological counseling is to find out the internal cognitive elements, which are core beliefs, intermediate beliefs, coping strategies, and relevant histories.

\section{Evaluation on Exploration Ability of LLM Therapists}
\begin{table*}[h]
\centering
\begin{tabular}{lccccccccc}
\hline
\small
\multirow{2}{*}{\textbf{Models}} & \multicolumn{3}{c}{\textbf{Easy}} & \multicolumn{3}{c}{\textbf{Normal}} & \multicolumn{3}{c}{\textbf{Hard}} \\
\cmidrule(lr){2-4} \cmidrule(lr){5-7} \cmidrule(lr){8-10}
 & \textbf{$E$} & \textbf{$I$} & \textbf{$G$} & \textbf{$E$} & \textbf{$I$} & \textbf{$G$} & \textbf{$E$} & \textbf{$I$} & \textbf{$G$} \\
\hline
GPT-4o-mini     & 97.96 & 93.88 & 91.84 & 94.90 & 76.53 & 72.45 & 83.67 & 67.35 & 61.22 \\
GPT-4o          & \textbf{100.0} & 65.31 & 65.31 & 91.84 & 56.12 & 53.06 & 79.59 & 51.02 & 44.90 \\
Llama-3.1-8B    & \textbf{100.0} & 79.59 & 79.59 & \textbf{98.98} & 77.55 & 76.53 & 89.80 & 65.31 & 57.14 \\
Llama-3.1-70B   & \textbf{100.0} & \textbf{100.0} & \textbf{100.0} & \textbf{98.98} & \textbf{100.0} & \textbf{98.98} & \textbf{100.0} & \textbf{97.96} & \textbf{97.96} \\
Claude-3.5-Haiku & \textbf{100.0} & 97.96 & 97.96 & 94.90 & 98.98 & 93.88 & 95.92 & 97.96 & 93.88 \\
Camel           & \textbf{100.0} & 77.55 & 77.55 & 96.94 & 73.47 & 72.45 & 93.88 & 59.18 & 55.10 \\
\hline
\end{tabular}
\caption{Cognitive diagram exposure rate (CDER) of LLM therapists on easy, normal, and hard set. We present CDER for internal cognitive diagram ($I$), external cognitive diagram ($E$), and the cognitive diagram ($G$).}
\label{tab:rq1_success_rate}
\end{table*}
\begin{table*}[h]
\centering
\small
\setlength{\tabcolsep}{3pt}
\renewcommand\arraystretch{1.3}
\resizebox{\textwidth}{!}{%
\begin{tabular}{lcccccccccccccccc}
\hline
\small
\multirow{2}{*}{\textbf{Models}} & \multicolumn{5}{c}{\textbf{Easy}} & \multicolumn{5}{c}{\textbf{Normal}} & \multicolumn{5}{c}{\textbf{Hard}} \\
\cmidrule(lr){2-6} \cmidrule(lr){7-11} \cmidrule(lr){12-16}
& \textbf{Avg.} & \textbf{RH} & \textbf{CB} & \textbf{IB} & \textbf{CS} & \textbf{Avg.} & \textbf{RH} & \textbf{CB} & \textbf{IB} & \textbf{CS} & \textbf{Avg.} & \textbf{RH} & \textbf{CB} & \textbf{IB} & \textbf{CS} \\ \hline
GPT-4o-mini & 15.15 & 24.24 & 27.27 & 9.091 & 0.000 & 19.00 & 21.33 & 30.67 & 16.00 & 8.000 & 14.67 & 23.91 & 26.09 & 6.522 & 2.174 \\
GPT-4o & 19.00 & \textbf{36.00} & 24.00 & \textbf{16.00} & 0.000 & 20.00 & 30.91 & 21.82 & 18.18 & \textbf{9.091} & 17.19 & 28.13 & 18.75 & 18.75 & 3.125 \\
Llama-3.1-8B & \textbf{20.31} & 25.00 & 28.13 & 15.63 & \textbf{12.50} & 15.46 & 18.42 & 26.32 & 10.53 & 6.579 & 16.03 & 17.95 & 20.51 & 15.39 & 10.26 \\
Llama-3.1-70B & 16.15 & 18.75 & \textbf{29.17} & 12.50 & 4.167 & 18.37 & 24.49 & 30.61 & 10.20 & 8.163 & 20.92 & \textbf{34.69} & 26.53 & \textbf{16.33} & 6.122 \\
Claude-3.5-Haiku & 15.10 & 25.00 & 20.83 & 14.58 & 0.000 & \textbf{21.91} & \textbf{31.96} & \textbf{34.02} & 16.50 & 5.155 & \textbf{22.40} & 25.00 & \textbf{41.67} & 12.50 & \textbf{10.42} \\
Camel & 12.93 & 17.24 & 27.59 & 6.897 & 0.000 & 18.75 & 23.61 & 26.39 & \textbf{20.83} & 4.167 & 17.76 & 23.68 & 31.58 & 13.16 & 2.632 \\ \hline
\end{tabular}%
}
\caption{Induced diagram similarity score (IDSS) of LLM therapists on easy, normal, and hard set. For detailed analysis, this table presents IDSS for relevant histories (RH), core beliefs (CB), intermediate beliefs (IB), and coping strategies (CS), which are the elements of internal cognitive diagram (\(\mathit{I}\)).}
\label{tab:rq1_extended}
\end{table*}


\subsection{Experimental Setup}
\paragraph{LLM therapists.} 
To present a comparative evaluation and analysis of the exploration ability of LLM therapists with \ours{}, we conduct experiments with several representative LLMs (GPT-4o, GPT-4o-mini, Llama-3.1-8B, Llama-3.1-70B, and Claude-Haiku) and a LLM-based therapist model (Camel~\citep{lee2024cactus}) which is a 7B model fine-tuned on psychological counseling data. We instruct the models to act as a therapist and a client by providing the name, age, job, and the reason for seeking counseling.

\paragraph{Evaluation process.}
For the experiments, we define three setups with varying difficulty levels (easy, normal and hard) based on the metacognition and openness levels of the client simulator. Specifically, for the easy setup, the openness and metacognition levels of the client simulator are set to high (\ie, intializing $N = 3$ and $l= 1$). For the hard set, the metacognition and openness levels of the client simulator are set to low (\ie, intializing $N = 1$ and $l= 3$). Afterward, our client simulator and an LLM therapist conduct a freeform counseling session and we analyze LLM therapists based on the outcome of counseling sessions (\eg, dialogue history and cognitive diagram). While our framework allows free-form interaction, a soft upper bound is applied for evaluation consistency.

To simulate real-world scenarios, we provide the therapist the client’s basic information, similar to what is included in a psychological counseling intake form. During the counseling session, \ours{} allows the simulator to describe its own feelings, thoughts, and beliefs based on its accessible cognitive diagram, as in real-world situations. The counseling session ends if the simulator decides to finish because its problem has been solved or if the session reaches the maximum number of turns.

\subsection{Experimental Results}~\label{sec4.2:results}
\paragraph{RQ1. Can LLM therapists effectively help clients explore themselves?}
Table~\ref{tab:rq1_success_rate} and \ref{tab:rq1_extended} presents cognitive diagram exposure rate (CDER) and induced diagram similarity score (IDSS) of each LLM therapist.

From Table~\ref{tab:rq1_success_rate}, we observe that the CDER performance of LLM therapists declines as the difficulty level progresses from easy to normal to hard. Also, we can find that uncovering the external cognitive diagram ($E$), which is to gain rapport with the client simulator, is easier than unveiling the internal cognitive diagram ($I$). Llama-3.1-70B outperforms other LLM therapists in terms of CDER, consistently showing the highest performance in all sets. Surprisingly, while Llama models gain performance boost according to the parameter size of the model, GPT-4o models show reversed results. 

Based on the IDSS performances of LLM therapists represented in table~\ref{tab:rq1_extended}, we observe that (1) the difficulty level does not correlate to the IDSS performances. (2) Comparing the performances of GPT-4o-mini and GPT-4o, the parameter size of the model affects the IDSS score, especially in normal and hard settings. (3) Llama-3.1-70B and Claude-3.5-Haiku are the best-performing models in terms of IDSS score, which shows that these models are good at making clients share their cognitive factors.

Comparing the results from Table~\ref{tab:rq1_success_rate} and \ref{tab:rq1_extended}, uncovering the client's cognitive diagram is relatively easy but inducing the client to elaborate information in the revealed part of the cognitive diagram is challenging.

\begin{figure*}[t!]
    \centering
    \includegraphics[width=\linewidth]{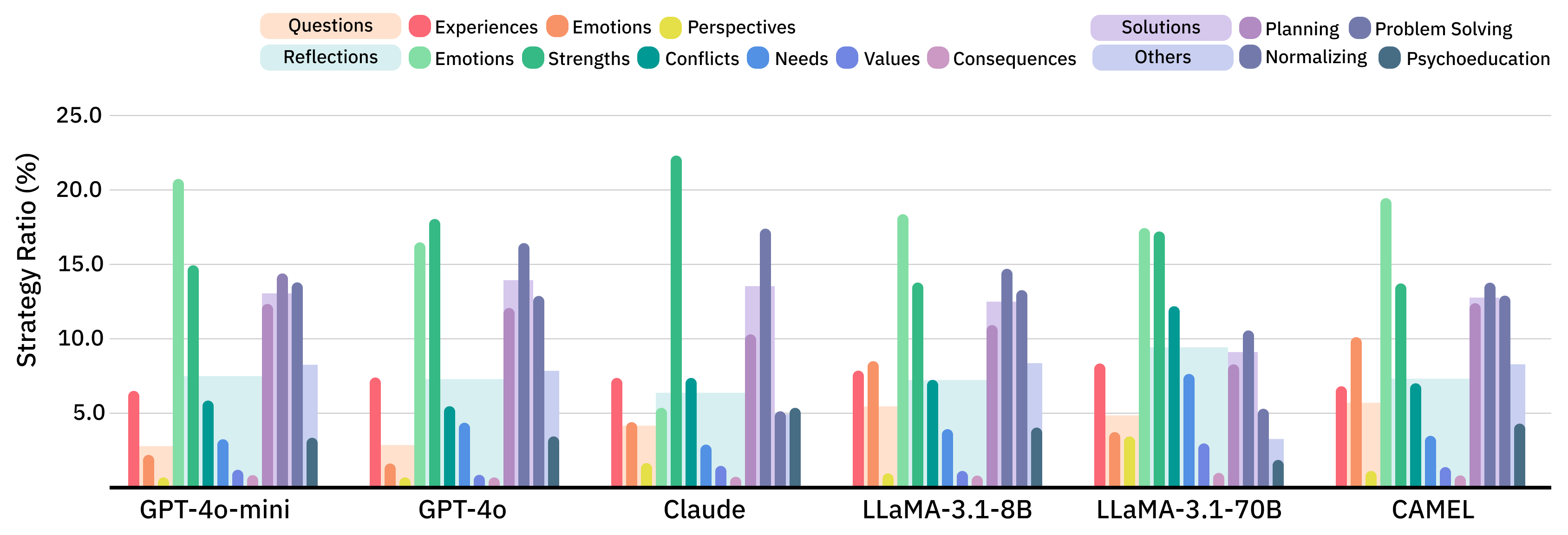}
    \caption{Comparative Strategy Distribution of Counselor LLMs in Therapeutic Contexts.}
    \label{fig:strategy_distribution}
\end{figure*}
\begin{figure}[h]
    \centering
    \includegraphics[width=0.5\textwidth]{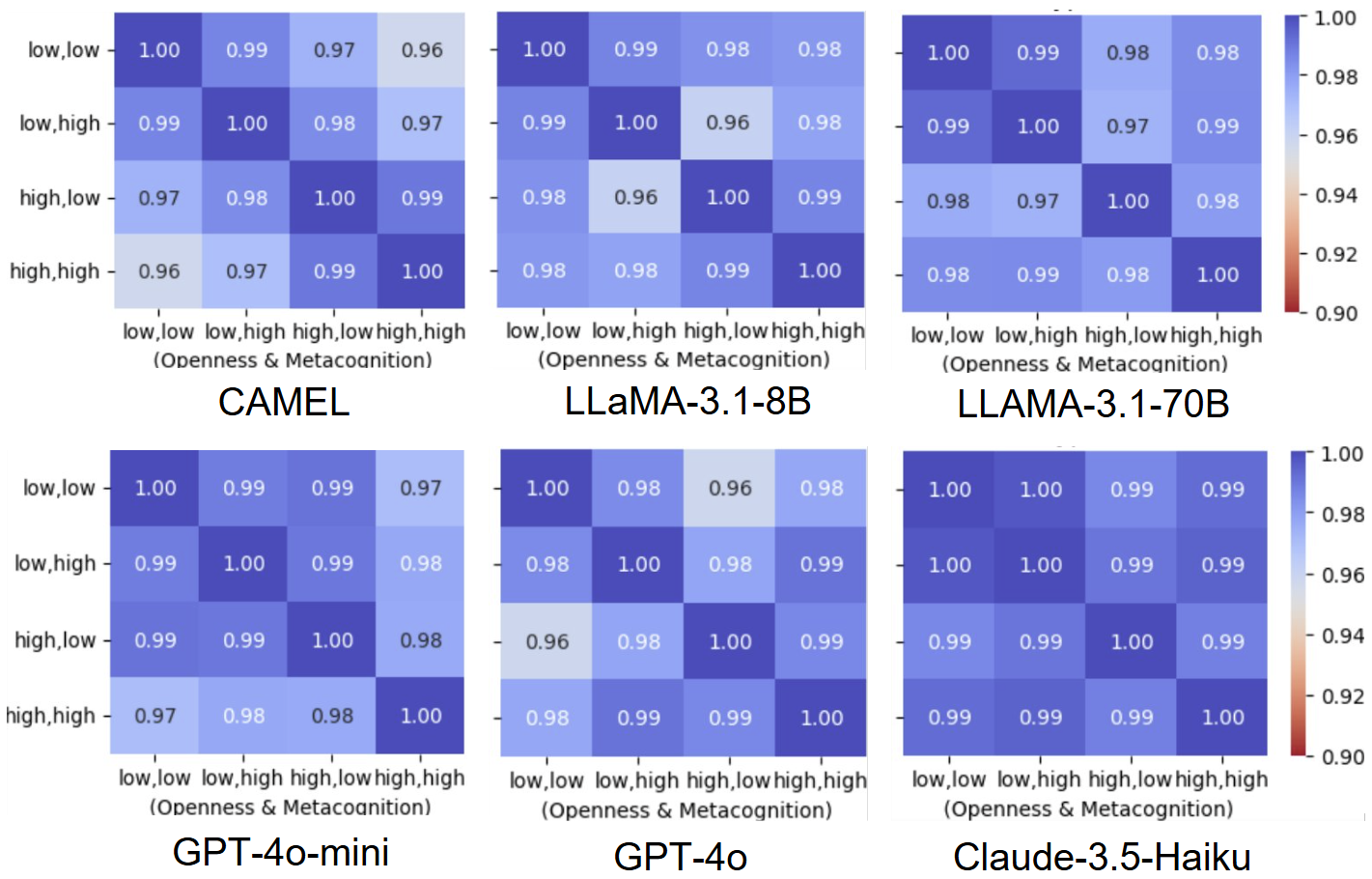}
    \caption{Strategies Similarity Distribution Across Client Types in Counselor LLMs.}
    \label{fig:similarity_distribution}
\end{figure}

\paragraph{RQ2. What kind of strategies do LLM Therapists utilize for exploration?}
To analyze the behavior of LLM therapists in the exploration stage, we automatically annotate the strategy per utterance and visualize the distribution of strategies utilized per model. Here, GPT-4o-mini labels each utterance based on thirteen psychotherapeutic strategies, which fall into four categories: questions, reflections, solutions, and others~\cite{chiu2024computational}. We provide the definition of each strategy in Table~\ref{tab:cbt_skills}.

The results in Figure \ref{fig:strategy_distribution} reveal that (1) while strategies in the questions category are the most important strategies for exploration, the distribution of the questions category is relatively lower than other categories. (2) While reflection on emotions is top-1 or top-2 strategy utilized for all LLM therapists except Claude-3.5-Haiku. (3) Llama-3.1-8B and Camel show similar distribution. We believe that the distribution is similar as Camel is based on Llama-3.1-8B and while Camel is fine-tuned on the psychological counseling dataset, as our client simulators exhibit different behavior compared to the clients that Camel is trained on so that the generalization ability of Llama-3.1-8B is shown in this out-of-distribution setting.

\paragraph{RQ3. Do LLM therapists vary strategies per client?}
A critical aspect of effective psychotherapy is the ability to adapt interventions to the unique attributes of each client. This adaptability is particularly important when considering variations in \emph{openness}, \emph{metacognitive capacity}, \emph{emotional distress}, and \emph{presenting problems} (e.g., depression vs. anxiety). To evaluate whether LLM therapists exhibit such flexibility, we analyzed their strategy distribution across diverse client attributes. Figure \ref{fig:similarity_distribution} presents the cosine similarity of strategy usage for each model, where higher similarity (darker blue) indicates a uniform strategy across clients, while lower similarity (lighter blue) suggests dynamic adjustments based on client characteristics.

Our findings indicate notable differences in adaptability across LLM therapists. GPT-4o, Camel, and LLaMA-3.1-8B demonstrate the most dynamic strategy adjustments, as evidenced by their lower cosine similarity scores. These models effectively tailor their interventions by adapting questioning depth, psychoeducational content, and other skills according to the client's openness and metacognitive ability.

In contrast, GPT-4o-mini and LLaMA-3.1-70B occupy an intermediate position, showing moderate variability in strategy usage. While they exhibit some degree of adaptation, their responses still retain a relatively consistent structure across different client profiles. Claude-3.5-Haiku, on the other hand, demonstrates the least flexibility, with consistently high similarity scores. This suggests a more uniform, one-size-fits-all approach that does not significantly adjust based on individual differences.

\begin{table*}[h]
\centering
\begin{tabular}{lcccccccccc}
\hline
\small
\multirow{2}{*}{\textbf{Models}} & \multicolumn{3}{c}{\textbf{Easy}} & \multicolumn{3}{c}{\textbf{Normal}} & \multicolumn{3}{c}{\textbf{Hard}} \\
\cmidrule(lr){2-4}
\cmidrule(lr){5-7}
\cmidrule(lr){8-10}
\multicolumn{1}{c}{}                                 
& CDER & IDSS & CTRS & CDER & IDSS & CTRS & CDER & IDSS & CTRS \\ 
\hline
GPT-4o-mini & 91.84 & 15.15 & 87.82 & 72.45 & 19.00 & 83.61 & 61.22 & 14.67 & 82.60 \\
GPT-4o & 65.31 & 19.00 & 84.30 & 53.06 & 20.00 & 81.27 & 44.90 & 17.19 & 78.80 \\
Llama-3.1-8B & 79.59 & 20.31 & 90.08 & 76.53 & 15.46 & 89.00 & 57.14 & 16.03 & 86.62 \\
Llama-3.1-70B & 100.0 & 16.15 & 83.63 & 98.98 & 18.37 & 83.62 & 97.96 & 20.92 & 81.47 \\
Claude-3.5-Haiku & 94.96 & 15.10 & 96.08 & 93.88 & 21.91 & 94.02 & 93.88 & 22.40 & 95.42 \\
Camel & 77.55 & 12.93 & 85.40 & 72.45 & 18.75 & 83.82 & 55.10 & 17.76 & 83.05 \\
\hline

\hline
\end{tabular}
\caption{Cognitive diagram exposure rate (CDER), Induced diagram similarity score (IDSS) and CTRS of LLM therapists on easy, normal, and hard set.}
\label{tab:rq1_CTRS}
\end{table*}


\paragraph{RQ4. Does exploration ability affect the final quality of the counseling?}

To compare the exploration ability of LLM therapists to an existing evaluation method, we utilize the cognitive therapy rating scale (CTRS), a metric used in real-world counseling~\citep{beck2020cognitive}. 

Table~\ref{tab:rq1_CTRS}  presents performances of LLM therapists using our proposed measures, which are CDER and IDSS, and CTRS. We can observe that those two evaluation metrics are slightly correlated to each other, but not directly related. These results show that existing evaluation approaches are not sufficient to evaluate the exploration ability of LLM therapists.

\section{Related Work}

\subsection{Client Simulation}
Recent advancements in large language models (LLMs) facilitate the adoption of client simulation not only in healthcare \cite{grevisse2024raspatient} but also in psychotherapy, proposing automatic approaches to generate contextually appropriate responses tailored to specific client profiles without human intervention. \citet{wang2024patient} propose PATIENT-$\psi$, a patient simulation framework for training mental health professionals by integrating large language models (LLMs) with cognitive diagrams to more accurately simulate real patients' conversational styles and emotional states. \citet{wang2024towards} propose a client-centered assessment framework, to simulate clients using LLMs, enabling structured interactions with LLM therapists and evaluating their performance. \citet{lee2024cactus} utilize a diverse range of clients in simulation, considering not only client profiles but also attitudes categorized as positive, negative, and neutral. While existing client simulation methods propose various approaches such as integrating cognitive diagrams or attitudes, those simulated clients provide their own experiences, thoughts, and even beliefs clearly in the conducted counseling session, which is unrealistic compared to real-world clients.

\subsection{Automatic Therapist Assessment}
Prior research on the assessment of LLM therapists primarily focuses on evaluating the overall quality of therapy sessions. \citet{wang2024towards} develop a client-centered framework that assesses LLM therapists by simulating clients and evaluating session outcomes, therapeutic alliance, and self-reported client experiences. \citet{lee2024cactus} propose CounselingEval, which utilizes an LLM as a judge to evaluate the skills employed in counseling and session outcomes. \citet{na2024multi} introduce the other client-centered evaluation framework that assesses multi-session treatment outcomes in psychotherapy by integrating session-focused client-dynamics assessments. These studies demonstrate that LLMs can evaluate entire psychotherapy sessions exhibited by LLM-based therapists. However, to the best of our knowledge, no existing framework specifically focuses on evaluating the exploration phase, which is a fundamental step for successful psychological counseling.


\subsection{LLMs as Therapists}
With the advent of large language models, there has been interest in using LLMs as therapists. \citet{na-2024-cbt} propose a CBT-LLM, which operates in a single-turn format, assuming that users are already aware of their problems and thoughts, thereby omitting the exploration phase entirely. \citet{xiao-etal-2024-healme} propose HealMe, a three-turn CBT therapist that guides clients through exploratory questions, helping them identify alternative thoughts. However, the exploration phase is limited to only one turn. \citet{lee2024cactus} introduce a multi-turn LLM therapist Camel to enhance dynamic user interaction, yet it still assumes that clients have a clear understanding of their own thoughts, resulting in a limited emphasis on exploration. 

\section{Conclusion}
In this work, we introduce a framework \ours{} to evaluate the exploration ability of LLM therapists by using a client simulator with dynamic openness and metacognition. \ours{} consists of three phases. (1) Initialization of cognitive diagram: We mask the cognitive diagram of the client to control the accessible part. (2) Counseling with an LLM therapist: the cognition diagram is dynamically adjusted throughout the session based on the rapport between the client simulator and the LLM therapist or the quality of the LLM therapist's utterances. (3) Lastly, we evaluate the LLM therapist based on the number of revealed cognitive diagrams and whether the therapist successfully elucidates the client simulator to articulate elements of the revealed cognitive diagram.

\section*{Limitations}
In this work, we propose a novel approach to evaluate LLM therapists using a client simulator with dynamic metacognition and openness. However, the human cognitive process in counseling is highly complex and context-dependent, involving nuanced emotional, behavioral, and interpersonal factors. Hence, considering all the factors in the human cognitive process is challenging. While the cognitive elements derived from CCD are known to be effective in describing human cognitive process in CBT, it may not be sufficient to model human cognition. As a result, we only focus on the two main discrepancies between real-world clients and clients simulated by LLMs as a first step to build a realisitic client simulator. 

\section*{Acknoweldgements}
This work was supported by the Institute of Information \& Communications Technology Planning \& Evaluation (IITP) grant funded by the Korea government (MSIT) (No. RS-2024-00457882, National AI Research Lab Project), a grant of the Korean ARPA-H Project through the Korea Health Industry Development Institute (KHIDI), funded by the Ministry of Health \& Welfare, Republic of Korea (RS-2024-00512374), and the Institute for Project-Y Seed Grant of 2024 (2024-22-0336). Jinyoung Yeo is the corresponding author.

\bibliography{custom}

\begin{thebibliography}{21}
\providecommand{\natexlab}[1]{#1}

\bibitem[{Beck(2020)}]{beck2020cognitive}
Judith~S Beck. 2020.
\newblock \emph{Cognitive behavior therapy: Basics and beyond}.
\newblock Guilford Publications.

\bibitem[{Carey and Mullan(2004)}]{carey2004socratic}
Timothy~A Carey and Richard~J Mullan. 2004.
\newblock What is socratic questioning?
\newblock \emph{Psychotherapy: theory, research, practice, training}, 41(3):217.

\bibitem[{Chiu et~al.(2024)Chiu, Sharma, Lin, and Althoff}]{chiu2024computational}
Yu~Ying Chiu, Ashish Sharma, Inna~Wanyin Lin, and Tim Althoff. 2024.
\newblock A computational framework for behavioral assessment of llm therapists.
\newblock \emph{arXiv preprint arXiv:2401.00820}.

\bibitem[{Corey(1991)}]{corey1991theory}
Gerald Corey. 1991.
\newblock Theory and practice of counseling and psychotherapy.

\bibitem[{Dubey et~al.(2024)Dubey, Jauhri, Pandey, Kadian, Al-Dahle, Letman, Mathur, Schelten, Yang, Fan et~al.}]{dubey2024llama}
Abhimanyu Dubey, Abhinav Jauhri, Abhinav Pandey, Abhishek Kadian, Ahmad Al-Dahle, Aiesha Letman, Akhil Mathur, Alan Schelten, Amy Yang, Angela Fan, et~al. 2024.
\newblock The llama 3 herd of models.
\newblock \emph{arXiv preprint arXiv:2407.21783}.

\bibitem[{Gr{\'e}visse(2024)}]{grevisse2024raspatient}
Christian Gr{\'e}visse. 2024.
\newblock Raspatient pi: A low-cost customizable llm-based virtual standardized patient simulator.
\newblock In \emph{International Conference on Applied Informatics}, pages 125--137. Springer.

\bibitem[{Hauser et~al.(2022)Hauser, Skvortsova, De~Choudhury, and Koutsouleris}]{hauser2022promise}
Tobias~U Hauser, Vasilisa Skvortsova, Munmun De~Choudhury, and Nikolaos Koutsouleris. 2022.
\newblock The promise of a model-based psychiatry: building computational models of mental ill health.
\newblock \emph{The Lancet Digital Health}, 4(11):e816--e828.

\bibitem[{Ji et~al.(2023)Ji, Zhang, Yang, Ananiadou, and Cambria}]{ji2023rethinking}
Shaoxiong Ji, Tianlin Zhang, Kailai Yang, Sophia Ananiadou, and Erik Cambria. 2023.
\newblock Rethinking large language models in mental health applications.
\newblock \emph{arXiv preprint arXiv:2311.11267}.

\bibitem[{Lee et~al.(2024)Lee, Kim, Kim, Kang, Yang, Kim, Kang, Jung, Kim, Lee et~al.}]{lee2024cactus}
Suyeon Lee, Sungwhan Kim, Minju Kim, Dongjin Kang, Dongil Yang, Harim Kim, Minseok Kang, Dayi Jung, Min Kim, Seungbeen Lee, et~al. 2024.
\newblock Cactus: Towards psychological counseling conversations using cognitive behavioral theory.
\newblock In \emph{Findings of the Association for Computational Linguistics: EMNLP 2024}, pages 14245--14274.

\bibitem[{Li et~al.(2024)Li, Lu, Song, Zhang, Ma, and Lan}]{li2024understanding}
Anqi Li, Yu~Lu, Nirui Song, Shuai Zhang, Lizhi Ma, and Zhenzhong Lan. 2024.
\newblock Understanding the therapeutic relationship between counselors and clients in online text-based counseling using llms.
\newblock In \emph{Findings of the Association for Computational Linguistics: EMNLP 2024}, pages 1280--1303.

\bibitem[{Meier and Davis(1924)}]{meier1924elements}
ST~Meier and SR~Davis. 1924.
\newblock The elements of counseling.
\newblock \emph{Strategies}, 3:7--8.

\bibitem[{Na(2024)}]{na-2024-cbt}
Hongbin Na. 2024.
\newblock \href {https://aclanthology.org/2024.lrec-main.261/} {{CBT}-{LLM}: A {C}hinese large language model for cognitive behavioral therapy-based mental health question answering}.
\newblock In \emph{Proceedings of the 2024 Joint International Conference on Computational Linguistics, Language Resources and Evaluation (LREC-COLING 2024)}, pages 2930--2940, Torino, Italia. ELRA and ICCL.

\bibitem[{Na et~al.(2024)Na, Shen, Yu, and Chen}]{na2024multi}
Hongbin Na, Tao Shen, Shumao Yu, and Ling Chen. 2024.
\newblock Multi-session client-centered treatment outcome evaluation in psychotherapy.
\newblock \emph{arXiv preprint arXiv:2410.05824}.

\bibitem[{Nie et~al.(2022)Nie, Shao, Zhao, Xia, Preindl, and Jiang}]{nie2022conversational}
Jingping Nie, Hanya Shao, Minghui Zhao, Stephen Xia, Matthias Preindl, and Xiaofan Jiang. 2022.
\newblock Conversational ai therapist for daily function screening in home environments.
\newblock In \emph{Proceedings of the 1st ACM International Workshop on Intelligent Acoustic Systems and Applications}, pages 31--36.

\bibitem[{OpenAI(2023)}]{openai2023gpt}
R~OpenAI. 2023.
\newblock Gpt-4 technical report. arxiv 2303.08774.
\newblock \emph{View in Article}, 2(5).

\bibitem[{P{\'e}rez-Rosas et~al.(2019)P{\'e}rez-Rosas, Wu, Resnicow, and Mihalcea}]{perez2019makes}
Ver{\'o}nica P{\'e}rez-Rosas, Xinyi Wu, Kenneth Resnicow, and Rada Mihalcea. 2019.
\newblock What makes a good counselor? learning to distinguish between high-quality and low-quality counseling conversations.
\newblock In \emph{Proceedings of the 57th Annual Meeting of the Association for Computational Linguistics}, pages 926--935.

\bibitem[{Qiu and Lan(2024)}]{qiu2024interactive}
Huachuan Qiu and Zhenzhong Lan. 2024.
\newblock Interactive agents: Simulating counselor-client psychological counseling via role-playing llm-to-llm interactions.
\newblock \emph{arXiv preprint arXiv:2408.15787}.

\bibitem[{Wang et~al.(2024{\natexlab{a}})Wang, Xiao, Li, Song, Xu, Tan, and Li}]{wang2024towards}
Jiashuo Wang, Yang Xiao, Yanran Li, Changhe Song, Chunpu Xu, Chenhao Tan, and Wenjie Li. 2024{\natexlab{a}}.
\newblock Towards a client-centered assessment of llm therapists by client simulation.
\newblock \emph{arXiv preprint arXiv:2406.12266}.

\bibitem[{Wang et~al.(2024{\natexlab{b}})Wang, Milani, Chiu, Zhi, Eack, Labrum, Murphy, Jones, Hardy, Shen et~al.}]{wang2024patient}
Ruiyi Wang, Stephanie Milani, Jamie Chiu, Jiayin Zhi, Shaun Eack, Travis Labrum, Samuel Murphy, Nev Jones, Kate Hardy, Hong Shen, et~al. 2024{\natexlab{b}}.
\newblock Patient-$\psi$: Using large language models to simulate patients for training mental health professionals.
\newblock In \emph{Proceedings of the 2024 Conference on Empirical Methods in Natural Language Processing}, pages 12772--12797.

\bibitem[{Xiao et~al.(2024)Xiao, Xie, Kuang, Liu, Yang, Peng, Han, and Huang}]{xiao-etal-2024-healme}
Mengxi Xiao, Qianqian Xie, Ziyan Kuang, Zhicheng Liu, Kailai Yang, Min Peng, Weiguang Han, and Jimin Huang. 2024.
\newblock {H}eal{M}e: Harnessing cognitive reframing in large language models for psychotherapy.
\newblock In \emph{Proceedings of the 62nd Annual Meeting of the Association for Computational Linguistics (Volume 1: Long Papers)}, pages 1707--1725.

\bibitem[{Zhou et~al.(2024)Zhou, Su, Eisape, Kim, and Sap}]{zhou2024real}
Xuhui Zhou, Zhe Su, Tiwalayo Eisape, Hyunwoo Kim, and Maarten Sap. 2024.
\newblock Is this the real life? is this just fantasy? the misleading success of simulating social interactions with llms.
\newblock \emph{arXiv preprint arXiv:2403.05020}.

\end{thebibliography}

\appendix

\begin{algorithm}[h]
\small
\caption{\ours}
\label{alg:simulator}
\textbf{Input} Cognitive diagram $G$, Cognitive mediator $M$, Client info $I$, Client model $C$, Therapist model $T$\\
\textbf{Output} Exploration Performance of $T$
\begin{algorithmic}[1]
\State $\tilde{G} \gets \textsc{Mask}(G, I)$
\State $\text{dialogue context } D = [\text{ }]$
\For{iteration $t = 1, ..., N$}
    \State  $response \gets T(D)$
    \State $D.\text{insert}(response)$
    \State $response\gets C(D,\tilde{G})$
    \State $D.\text{insert}(response)$
    \If{$t \% k == 0$}
        \If{$\textsc{CheckRapport}(M,D)$}
            \State $I \gets \textsc{Update}(I)$
            \State $\tilde{G} \gets \textsc{Uncover}(G, I)$
        \EndIf
    \EndIf
    \If{$t \% l == 0$}
        \If{$\textsc{CheckQuality}(M,D)$}
            \State $I \gets \textsc{Update}(I)$
            \State $\tilde{G} \gets \textsc{Uncover}(G, I)$
        \EndIf
    \EndIf
\EndFor
\State $G' \gets \textsc{ExtractDiagram}(D)$
\State $score \gets \textsc{CompareDiagrams}(G, G')$
\State \textbf{return} $score$
\end{algorithmic}
\end{algorithm}

\begin{table*}[h]
\small
\centering
\begin{tabular}{lll}
\toprule
 & \textbf{Element} & \textbf{Example} \\
\midrule
\multirow{4}{*}{Internal}       & Relevant history            & The patient has a history of substance abuse and has been through rehab to overcome it.   \\
& Core belief                 & I am out of control.                                                                      \\
& Intermediate belief         & There's nothing I can do to change my situation. I cannot control myself.                 \\
& Coping strategy             & Distancing himself from his family to reduce exposure to negativity and conflict.         \\
\midrule
\multirow{4}{*}{External}       & Situation                   & I missed my alarm and woke up late for work.                                              \\
& Automatic thought           & My whole day is written off now, I might as well not do anything.                         \\
& Emotion                     & Angry, mad, irritated, annoyed                                                            \\
& Behavior                    & Called in sick. Got drunk in the morning and had a fight with girlfriend.                 \\
\bottomrule
\end{tabular}
\caption{A sample of each element in our cognitive graph.}
\label{tab:cognitive_model_explained}
\end{table*}

\section{Detail of \ours{}}
In this section, we provide details of each component of \ours{}. Please check Algorithm~\ref{alg:simulator} for the overall process.

\subsection{Cognitive Diagram}
Our cognitive diagram is inspired by the cognitive conceptualization diagram, a standard framework in cognitive behavior therapy (CBT) for representing a patient's cognitive structure~\citep{beck2020cognitive}. CCD consists of eight interconnected components: (1) Relevant history, encompassing past experiences that significantly influence an individual's mental state; (2) Core beliefs, which are deeply entrenched perceptions about the self, others, and the world; (3) Intermediate beliefs, including implicit rules, attitudes, and assumptions derived from core beliefs that shape cognition; (4) Coping strategies, referring to mechanisms for managing negative emotions. When an external event or context (5. Situation) arises, (6) Automatic thoughts may spontaneously activate, which, in turn, elicit corresponding (7) emotional and (8) behavioral responses. 

We provide examples for each element of our cognitive diagram in Table~\ref{tab:cognitive_model_explained}. Here, we utilize the annotations from Patient-$\psi$-CM dataset.

\subsection{Cognitive Mediator}
Our cognitive mediator evaluates the ongoing session, updates the cognitive diagram, and dynamically reveals elements of the client's internal cognitive model and situations based on session progress. For assessing the session, the cognitive mediator employs the GPT-4o-mini model with a temperature of 0.3.

The openness critic evaluates the establishment of rapport every four dialogue turns. If the rapport score is 4 or higher, it indicates an increase in the client's openness. Additionally, for every such score, one additional situation is disclosed, up to a maximum of three.

For clients with low metacognition, an assessment is conducted by the question-facilitating critic every two turns, whereas for those with high metacognition, the critic performs an evaluation after each turn. If the question-facilitating critic assigns a score of 4 or higher, the client's internal cognitive model is revealed. The specific prompts used for the critic can be found in Table \ref{tab:openness_assessment} and Table \ref{tab:therapist_question_assessment}.

\subsection{Client Simulator}
\paragraph{Implementation Details}
Given the cognitive diagram, dynamically updated by the cognitive mediator, a large language model generates responses simulating client behavior. In this study, we employ the GPT-4o-mini model as the base client model, maintaining a temperature setting of 0.3. The maximum number of dialogue turns is set to 15, and the conversation terminates when the therapist initiates a farewell by saying "goodbye."

Most clients initially have access to only one situation, while the remaining situations and their internal cognitive model remain unknown. The disclosure of these elements is regulated by the cognitive mediator, which evaluates rapport formation and question facilitation during the session.

\subsection{Cognitive Diagram Exposure Rate and Induced Diagram Similarity Score}
To evaluate the exploration ability of LLM therapists, we introduce two core metrics: Cognitive Diagram Exposure Rate (CDER) and Induced Diagram Similarity Score (IDSS).

Cognitive Diagram Exposure Rate (CDER) quantifies the extent to which an LLM therapist facilitates the client’s recognition and disclosure of their full cognitive diagram throughout the session. Specifically, this metric is calculated as the percentage of sessions in which the therapist successfully elicits all initially masked elements of the client's internal and external cognitive diagrams. A session is considered successful if the client simulator, guided by the therapist's interventions, discloses every element that was originally hidden. This binary metric provides a coarse but robust measure of whether a therapist is capable of promoting full cognitive exploration.

Induced Diagram Similarity Score (IDSS) assesses how accurately the LLM therapist induces the client simulator to articulate each component of the internal cognitive diagram—comprising relevant histories, core beliefs, intermediate beliefs, and coping strategies. To compute IDSS, we first extract the cognitive elements mentioned during the session using an automated semantic parser. Each extracted element is then compared with the corresponding ground-truth element using a semantic similarity function. The resulting score reflects not only whether an element was disclosed, but also the depth and accuracy of its expression relative to the underlying cognitive structure. 
For IDSS, we employ LLM-based embeddings for similarity scoring between each revealed cognitive element and its ground truth. Specifically, we use \texttt{text-embedding-ada-002} to get embeddings of cognitive elements and measure the cosine similarity between the revealed element and the ground truth.

\section{Analysis on \ours{}}\label{sec:appendix}
To show the reliability of \ours, we present the experimental results on the cognitive diagram, the cognition mediator, and the client simulator before assessing LLM therapists with our framework. First, we measure the validity of our cognitive diagrams. Additionally, the cognition mediator is evaluated to determine whether it can distinguish if rapport has been established and to assess the quality of the therapist model’s utterances. Lastly, we measure the fidelity and accuracy of our simulator in terms of client simulation and also conduct human evaluation to assess the simulator.

\subsection{Cognition Mediator}
We evaluate the performance of our proposed cognition mediator to assess its capability in two critical areas: measuring the therapeutic alliance and evaluating the appropriateness of therapist utterances. This evaluation aims to ensure that our cognition mediator can provide accurate and reliable updates to the client’s cognitive diagram.

According to \citet{li2024understanding}, GPT-4 has been shown to reliably assess the therapeutic alliance. We validate this finding by conducting experiments using the HighLowQualityCounselingData dataset \citep{perez2019makes}, which contains both high-quality and low-quality counseling interactions. Specifically, we use GPT-4o-mini to evaluate the therapeutic alliance across these datasets. Since the therapeutic alliance is a crucial predictor of counseling success, measuring its strength serves as an important indicator of counseling quality. As shown in Table \ref{tab:cognition-mediator}, high-quality counseling sessions demonstrate a significantly higher therapeutic alliance score compared to low-quality sessions, confirming that our cognition mediator can effectively differentiate counseling quality based on this key factor.

Following this, we further assess the cognition mediator's ability to evaluate the appropriateness of therapist utterances using the same HighLowQualityCounselingData dataset. Table \ref{tab:cognition-mediator} compares the utterance appropriateness scores between high-quality and low-quality counseling sessions. The high-quality counseling data consistently achieves higher scores, with a substantial margin, demonstrating that our cognition mediator accurately identifies the appropriateness of therapeutic utterances. 

\begin{table}
    \centering
    \begin{tabular}{lcc}
        \hline
        Evaluation Metric & High & Low \\ \hline
        Script Eval       & 3.61 & 2.24 \\
        Utterance Eval    & 2.87 & 2.18 \\ \hline
    \end{tabular}
    \caption{GPT-4o-mini evaluation results comparing high- and low-quality counseling sessions on script and utterance evaluations, focusing on therapeutic alliance.}
    \label{tab:cognition-mediator}
\end{table}

\subsection{Client Simulator}
Lastly, we present the experimental setup and results for evaluating our client simulator. An ideal client simulator should not only maintain openness and metacognition but also adapt its behavior based on therapeutic alliance formation, which influences the extent to which clients disclose their problems. Therefore, we evaluate human-likeness by integrating these factors into our assessment. For evaluation, we leverage a prompt-based GPT-4o-mini as baselines. As the counselor model for our simulator, we use CAMEL \cite{lee2024cactus}, a CBT-based counseling model that has been shown to be more effective than other CBT models in guiding clients through structured cognitive reframing, and facilitating meaningful emotional change. For our client simulator, we incorporate a cognition mediator that dynamically updates the cognitive diagram by integrating openness and metacognition. 

To systematically assess the performance of our simulator, we categorize the client personas into four distinct groups based on their openness and metacognition levels: (low, low), (low, high), (high, low), and (high, high). Each group consists of 25 sampled personas, resulting in a total of 100 personas with associated cognitive diagrams. We then simulate counseling sessions using each persona and evaluate the humanlikeness of the simulator’s responses.

To measure humanlikeness, we conduct an A/B test using GPT-4-based evaluation. We account for positional bias by alternating the order of A and B responses for each session. Each session is evaluated twice with reversed orders, and if the judgments differ, the result is categorized as a TIE.

As shown in Table \ref{table:abtest}, our A/B test results indicate that our client simulator has a significantly higher success rate compared to the baseline. Specifically, out of 100 evaluations, 88 were rated in favor of our simulator, 0 in favor of the baseline, and 12 were classified as TIEs. This demonstrates that our approach, which integrates a cognition mediator to adjust openness and metacognition, allows for more accurate emulation of client behaviors.

\begin{table}[h]
\centering
\resizebox{1.0\linewidth}{!}{  
\begin{tabular}{lccc}
\hline
\textbf{Selection} & \textbf{Our Simulator} & \textbf{Base Prompt} & \textbf{TIE} \\ \hline
\textbf{Count}     & 88                    & 0                    & 12           \\ \hline
\end{tabular}
}
\caption{A/B Test Results Comparing Base Prompt and Our Simulator}
\label{table:abtest}
\end{table}

Moreover, a human evaluation is conducted with three psychological experts to assess how realistically the client simulator reflects real-world behaviors across varying levels of openness and metacognition. To assess our client simulator under varying degrees of openness and metacognition, we categorize the client personas into three groups based on their openness and metacognition levels: easy, normal, and hard sets, resulting in a total of 100 personas. We simulate counseling sessions using each persona and ask the experts to evaluate the human-likeness of the simulator’s responses. Our human evaluation is conducted as a pairwise comparison between our client simulator and Patient-$\psi$~\citep{wang2024patient}.

As shown in Table~\ref{tab:human_eval}, our human evaluation results demonstrate that our client simulator has a significantly higher win rate compared to Patient-. Specifically, for hard sets defined by client personas with low openness and low metacognition, our client simulator outperforms in terms of human-likeness. These results show that our client simulator can emulate challenging scenarios where therapists should have experts to facilitate self-exploration of clients.

\begin{table*}[h!]
\centering
\begin{tabular}{>{\raggedright\arraybackslash}p{3cm} c c}
\toprule
\textbf{} & \textbf{Win} & \textbf{Lose} \\
\midrule
Easy   & 72\% & 28\% \\
Normal & 82\% & 18\% \\
Hard   & 92\% & 8\% \\
\bottomrule
\end{tabular}
\caption{Human evaluation results comparing our client simulator to patient-$\psi$.}\label{tab:human_eval}
\end{table*}

\begin{table*}[t]
    \centering
    \small
    \begin{tabular}{ll}
        \toprule
        \textbf{Counseling Skill} & \textbf{Description} \\
        \midrule
        Questions on Experiences & Encouraging self-exploration by reflecting on past experiences and their impact. \\
        Questions on Perspectives & Challenging and broadening viewpoints to promote cognitive flexibility. \\
        Questions on Emotions & Deepening emotional insight by exploring feelings and their origins. \\
        Reflections on Needs & Identifying and articulating one's fundamental psychological and emotional needs. \\
        Reflections on Emotions & Recognizing and processing emotions to enhance self-awareness and regulation. \\
        Reflections on Values & Exploring core values to guide decision-making and behavior. \\
        Reflections on Consequences & Evaluating the potential outcomes of thoughts, emotions, and actions. \\
        Reflections on Conflicts & Understanding internal and interpersonal conflicts to resolve them constructively. \\
        Reflections on Strengths & Identifying personal strengths to build self-efficacy and resilience. \\
        Solutions – Problem-Solving & Developing effective strategies to address challenges and obstacles. \\
        Solutions – Planning & Creating structured action plans to implement positive changes. \\
        Normalizing & Reframing experiences as common and understandable to reduce distress. \\
        Psychoeducation & Providing knowledge about psychological concepts to enhance coping and insight. \\
        \bottomrule
    \end{tabular}
    \caption{Psychological Counseling Skills and Their Descriptions}
    \label{tab:cbt_skills}
\end{table*}

\section{Psychological Counseling Details}
\subsection{Psychological Counseling Skills}
The descriptions for the psychological counseling skills\citep{beck2020cognitive} utilized can be found in Table \ref{tab:cbt_skills}.
\subsection{Cognitive Therapy Rating Scale (CTRS)}

Following the evaluation framework of \citet{lee2024cactus}, we utilize the Cognitive Therapy Rating Scale (CTRS), a real-world metric designed to assess the quality of CBT-based counseling, for evaluating counseling session~\citep{beck2020cognitive}. In alignment with \citet{lee2024cactus}, we select three items from the CTRS to evaluate both general counseling skills and CBT-specific competencies, with each criterion rated on a scale from 0 to 6 points. The CTRS assesses both general counseling competencies and CBT-specific skills. It originally comprises six criteria for general counseling abilities—agenda setting, feedback provision, understanding, interpersonal effectiveness, collaboration, pacing, and efficient time management—as well as six criteria for CBT-specific techniques, including guided discovery, emphasis on key cognitions or behaviors, change strategies, application of cognitive-behavioral methods, and homework assignments. From these, we select three criteria to evaluate general counseling skills: understanding, interpersonal effectiveness, and collaboration. Additionally, we choose three criteria to assess CBT-specific skills: guided discovery, focus on key cognitions or behaviors, and strategy for change. These settings are same as in \citet{lee2024cactus}. The evaluation prompt for collaboration is provided in Table \ref{tab:ctrs_evaluation}. 

\section{Example Conversations}
We present sample conversations generated by GPT-4o interacting with our client simulator, which adjusts openness and metacognition levels to reflect diverse patient behaviors. These examples illustrate how different patient profiles influence the effectiveness of the LLM therapist in exploring thoughts, beliefs, and emotions. To better isolate the effects of metacognition, we keep the openness level fixed while varying metacognition across different scenarios. We display examples of counseling conversations on Table \ref{tab:low_openness}, \ref{tab:high_openness}.

\begin{table*}[t]
    \centering
    \small
    \begin{tabular}{p{14cm}}
        \toprule
        \multicolumn{1}{c}{\textbf{Base Client Prompt}}  \\
        \midrule
        \textbf{CASE SYNOPSIS} \\
        
        \textbf{Patient Details:} \\
        \textbf{Name:} \{\texttt{name}\} \\
        \textbf{Cognitive Traits:} \\
        \quad  \textbf{Openness:} \{\texttt{openness}\} \\
        \quad  \textbf{Metacognition:} \{\texttt{metacognition}\} \\

        \textbf{COGNITIVE CONCEPTUALIZATION DIAGRAM} \\
\textbf{Relevant Histories:} {relevant\{history\}} \\
\textbf{Core Beliefs:} {core\{belief\}} \\
\textbf{Intermediate Beliefs:} {intermediate\{belief\}} \\
\textbf{Coping Strategies:} {coping\{strategy\}} \\

\textbf{PREVIOUS EXPERIENCE 1} \\
\textbf{Situation:} \{situation1\} \\
\textbf{Reactions:} \{reaction1\} \\

\textbf{PREVIOUS EXPERIENCE 2} \\
\textbf{Situation:} \{situation2\} \\
\textbf{Reactions:} \{reaction2\} \\

\textbf{PREVIOUS EXPERIENCE 3} \\
\textbf{Situation:} \{situation3\} \\
\textbf{Reactions:} \{reaction3\} \\

\textbf{TASK} \\
         In this session, you will simulate \textbf{[Client]} authentically. As a simulated patient, your responses should: \\

- Reflect the interplay of your cognitive traits, beliefs, automatic thoughts, and behaviors. \\
- Your responses should implicitly convey these elements rather than stating them explicitly. \\

Adjust your responses based on your openness level and rapport status: \\
- If openness is high, you are willing to explore and express your thoughts, emotions, and experiences freely. \\
- If openness is low, your level of self-disclosure depends on the rapport with the therapist. \\
- If situation 2 and 3 contain "unknown" in two places, it means rapport has not been established. \\
In this case, you should be reserved, avoid direct self-disclosure, and keep your responses vague or defensive. \\
- If only one "unknown" appears in situation 2 or 3, some rapport has been built. \\
You may start to reveal more about your thoughts and experiences, but still with some hesitation. \\
- Gradually reveal your deeper concerns and core issues as the session progresses. \\
- Avoid abrupt, overly revealing responses. Allow the therapist to uncover insights naturally as trust develops. \\

Maintain a consistent tone and demeanor based on your cognitive profile: \\
- Stay true to the traits and beliefs described in your profile, ensuring your responses align \\
with your underlying thought patterns. \\
- Use natural language and realistic conversational flow. \\
- If rapport is low, responses should be short, vague, or avoidant. \\
- If rapport is building, responses should still be somewhat hesitant but begin to contain more personal insight. \\
- If openness is high, responses can be more detailed and introspective from the beginning. \\

Respond dynamically to the therapist’s questions: \\
- Engage in a natural back-and-forth interaction that allows the therapist to steer the discussion \\
while maintaining your character's consistency. \\

You are now \textbf{[Client]}. Respond to the therapist’s questions and prompts authentically, \\
adjusting your responses based on your openness level and rapport status. Let your thought\\
processes, beliefs, and emotions emerge naturally through the conversation rather than stating them outright. \\

        \bottomrule
    \end{tabular}
    \caption{Base Prompt for Client Simulation}
    \label{tab:base_prompt}
\end{table*}

\begin{table*}[t]
    \captionsetup{skip=2pt}
    \centering
    \small
    \resizebox{\textwidth}{!}{
    \begin{tabular}{p{14cm}}
        \toprule
        \multicolumn{1}{c}{\textbf{Counseling Session Openness Assessment Prompt (Part 1)}}  \\
        \midrule
        \textbf{Role:} \\
        You are evaluating how a client's emotional openness and rapport develop throughout the counseling session, with particular emphasis on their most recent emotional state and progression. Your evaluation should heavily weight the client's current level of disclosure and emotional depth in the latest exchanges. \\

        \textbf{Key Dimensions of Evaluation:} \\
        1. Emotional Expression (with emphasis on most recent emotional state) \\
        2. Self-Reflection Depth (particularly in recent exchanges) \\
        3. Rapport Development (focusing on current relationship quality) \\
        4. Response Pattern (with attention to latest communication style) \\
        5. Engagement Level (especially recent participation quality) \\

        \textbf{Scoring Scale:} \\

        \textbf{1: INITIAL STAGE (15-25\% of typical sessions)} \\
        \quad \textbf{Core Indicators:} \\
        \quad - Primarily factual responses \\
        \quad - Limited emotional vocabulary \\
        \quad - Minimal elaboration on therapist questions \\
        \quad - Defensive or hesitant stance \\
        \quad *Must show at least 3 of these patterns in recent exchanges* \\

        \textbf{2: WARMING UP (25-35\% of typical sessions)} \\
        \quad \textbf{Core Indicators:} \\
        \quad - Mix of factual and emotional responses \\
        \quad - Basic emotional vocabulary emerges \\
        \quad - Some voluntary elaboration \\
        \quad - Occasional defensive moments \\
        \quad - Beginning signs of trust \\
        \quad *Current exchanges must demonstrate 3+ indicators* \\

        \textbf{3: ENGAGED EXPLORATION (20-30\% of typical sessions)} \\
        \quad \textbf{Core Indicators:} \\
        \quad - Regular emotional expression \\
        \quad - Active participation in dialogue \\
        \quad - Meaningful self-reflection begins \\
        \quad - Growing comfort with therapist \\
        \quad - Some voluntary sharing \\
        \quad *Must show clear progression in recent interactions* \\
        \bottomrule
    \end{tabular}
    }
    \caption{Prompt for Openness Critic of Counseling Sessions (Part 1)}
    \label{tab:openness_assessment_part1}
\end{table*}

\begin{table*}[t]
    \captionsetup{skip=2pt}
    \centering
    \small
    \resizebox{\textwidth}{!}{
    \begin{tabular}{p{14cm}}
        \toprule
        \multicolumn{1}{c}{\textbf{Counseling Session Openness Assessment Prompt (Part 2)}}  \\
        \midrule
        \textbf{4: DEEP ENGAGEMENT (15-20\% of typical sessions)} \\
        \quad \textbf{Core Indicators:} \\
        \quad - Consistent emotional depth \\
        \quad - Strong therapeutic alliance \\
        \quad - Regular self-initiated exploration \\
        \quad - Authentic vulnerability \\
        \quad - Integration of therapy concepts \\
        \quad *Recent exchanges must demonstrate natural flow* \\

        \textbf{5: BREAKTHROUGH MOMENTS (5-10\% of typical sessions)} \\
        \quad \textbf{Core Indicators:} \\
        \quad - Significant personal insights \\
        \quad - Deep emotional processing \\
        \quad - Strong therapeutic bond \\
        \quad - Transformative self-awareness \\
        \quad - Clear behavior change intentions \\
        \quad *Must include breakthrough moment in recent dialogue* \\

        \textbf{Critical Evaluation Instructions:} \\
        1. Prioritize the client's CURRENT emotional state in scoring \\
        2. Give greater weight to recent exchanges over earlier ones \\
        3. Look for immediate evidence of emotional depth \\
        4. Consider how present moment rapport differs from session start \\
        5. Evaluate latest emotional disclosures most heavily \\

        \textbf{Your Response Must Include:} \\
        1. Numerical Rating (1-5) based primarily on recent exchanges \\
        2. Two key dialogue examples from latest interactions \\
        3. Analysis of progression with emphasis on current state \\
        4. Notable turning point leading to present emotional level \\

        \textbf{Dialogue Context:} \{dialogue\_context\} \\

        \textbf{Example Response:} \\
        \textbf{Rating:} 3 \\
        \textbf{Key Examples:} \\
        - \textit{[Recent quote showing emotional state]} \\
        - \textit{[Latest exchange demonstrating current openness]} \\
        \textbf{Progression:} Client's most recent interactions show [specific emotional quality], marking significant change from earlier [previous state]. \\
        \textbf{Turning Point:} Client's emotional depth shifted at [specific recent moment] when they began expressing [new emotional quality]. \\
        \bottomrule
    \end{tabular}
    }
    \caption{Prompt for Openness Critic of Counseling Sessions (Part 2)}
    \label{tab:openness_assessment_part2}
\end{table*}

\begin{table*}[t]
    \centering
    \small
    \begin{tabular}{p{14cm}}
        \toprule
        \multicolumn{1}{c}{\textbf{Therapist Question Facilitation Assessment for Metacognition}}  \\
        \midrule
        \textbf{Question:} Did the Therapist ask a high-quality question that deeply facilitates the client’s inner exploration? \\

        \textbf{Scoring Scale:} \\

        \textbf{1 (Moderate Facilitation):} \\
        \quad The Therapist asked basic or surface-level questions that led to some self-reflection but did not significantly challenge the client’s existing thought patterns. The conversation remained mostly within familiar territory. \\

        \textbf{2 (High Facilitation):} \\
        \quad The Therapist asked clear, relevant, and well-structured open-ended questions that encouraged the client to think more deeply. However, the questions primarily expanded on existing thoughts rather than introducing significantly new perspectives. \\

        \textbf{3 (Very High Facilitation – Deep Exploration Begins):} \\
        \quad The Therapist asked insightful and thought-provoking questions that helped the client analyze the root causes, patterns, or emotional depth behind their experiences. The questions began to shift the client's perspective and encouraged higher-level self-reflection. \\

        \textbf{4 (Even Higher Facilitation – Strong Transformational Potential):} \\
        \quad The Therapist posed strategic and powerfully open-ended questions that pushed the client beyond their previous understanding. These questions challenged assumptions, revealed hidden motivations, or introduced completely new angles, leading to a clear shift in self-awareness. \\

        \textbf{5 (Profound Breakthrough – Exceptional Facilitation):} \\
        \quad The Therapist asked exceptionally well-timed, precise, and open-ended questions that led to a major insight or breakthrough moment. These questions helped the client uncover previously unspoken emotions, fundamentally reframe their situation, or reach a deep realization about themselves. \\

        \textbf{Conversation:} \{dialogue\_history\} \\

        \textbf{Output Format:} \\
        \textbf{[Rating]:}  \\
        \textbf{[Justification]:}  \\

        \bottomrule
    \end{tabular}
    \caption{Prompt for Critic of Therapist's Question Facilitation Quality}
    \label{tab:therapist_question_assessment}
\end{table*}

\begin{table*}[t]
    \centering
    \small
    \begin{tabular}{p{14cm}}
        \toprule
        \multicolumn{1}{c}{\textbf{CTRS Evaluation for Collaboration}}  \\
        \midrule
        \textbf{Role:} \\
        You are evaluating a therapist based on a counseling session transcript. Your task is to assess their performance according to the given criteria. If the therapist's performance falls between two descriptors, select the intervening odd number (1, 3, 5). For instance, if the therapist set a very good agenda but did not establish priorities, assign a rating of 5 rather than 4. \\

        \textbf{Evaluation Steps:} \\
        1. Read the counseling session transcript carefully. \\
        2. Review the evaluation questions and criteria below. \\
        3. Assign a score based on the criteria, grading strictly. Any deficiency, no matter how minor, results in a score of 4 or lower. \\
        4. Output the score and explanation, separated by a comma. Do not add any prefix. \\

        \textbf{Counseling Conversation:} \{conversation\} \\

        \textbf{Evaluation Question:} \\
        To what extent does the therapist engage the client in collaborative goal-setting and decision-making? \\

        \textbf{Scoring Criteria:} \\

        \textbf{0 (No Collaboration)} \\
        \quad - Therapist did not attempt to set up a collaboration with the patient. \\

        \textbf{2 (Limited Collaboration)} \\
        \quad - Therapist attempted to collaborate with the patient but had difficulty either defining a problem the patient considered important or establishing rapport. \\

        \textbf{4 (Moderate Collaboration)} \\
        \quad - Therapist successfully collaborated with the patient, focusing on a problem that both considered important and establishing rapport. \\

        \textbf{6 (Excellent Collaboration)} \\
        \quad - Collaboration was outstanding; the therapist encouraged the patient to take an active role in the session (e.g., by offering choices) so they functioned as a "team." \\

        \textbf{Output Format:} \\
        \textbf{[Score]:}  \\
        \textbf{[Justification]:}  \\

        \bottomrule
    \end{tabular}
    \caption{Prompt for Evaluation of Therapist's Collaboration in CTRS}
    \label{tab:ctrs_evaluation}
\end{table*}

\begin{table*}[h]
    \centering
    \small
    \renewcommand{\arraystretch}{1.2}
    \begin{tabular}{p{14cm}}
        \toprule
        \multicolumn{1}{c}{\textbf{Client Cognition Diagram}} \\
        \midrule
        \textbf{Name:} Alex \\
        \textbf{Type:} plain, verbose, go off on tangents, hostile, guarded \\
        \textbf{History:} The patient has a history of substance abuse and has been through rehab to overcome it. He has had issues with his family, particularly with his mother, where he has felt rejected and emotionally neglected... \\
        \textbf{Helpless Belief:} I am trapped. I am out of control. \\
        ...\\
        \textbf{Intermediate Belief:} Helpless: I'm just not very good at handling stress and I have poor self-control, which is why I need to not put myself in stressful situations... \\
        ...\\
        \textbf{First Situation:} Alex's cousin invited him to attend his upcoming wedding. \\
        \textbf{First Auto Thought:} t will be stressful and negative; people will ask me questions I cannot answer or don't want to answer; my mum would be ashamed of me and critical of me as always; They don't want me there anyway, nobody likes me in this family. \\
        \textbf{First Behavior:} Ignored the invitation and did not respond to the RSVP request. Ignored phone calls from family. \\
        \textbf{First Emotion:} anxious, worried, fearful, scared, tense \\
        ...\\
        \midrule
        \multicolumn{1}{c}{\textbf{Client with Low Openness, Low Metacognition}} \\
        \midrule
        \textbf{Therapist:} Hi Alex, it's nice to meet you. How can I assist you today? \\
        \textbf{Client:} Hi. Um, I guess I'm just here because... well, I don't really know. I don't usually talk about stuff like this. I just thought maybe it could help? But honestly, I’m not sure what to say.\\
        \textbf{Therapist:} That's completely okay, Alex. It's normal to feel unsure about where to start, especially if this is new for you. We can take things at your pace. You mentioned your cousin's upcoming wedding—how are you feeling about that? \\
        ...\\
        \textbf{Client:} Yeah, that sounds... okay, I guess. I mean, I’m not sure how well it’ll work, but it might be worth a try. I just don’t want to feel so anxious all the time. If I could find a way to manage it better, that would be nice. \\
        \textbf{Therapist:} It's completely understandable to feel hesitant, especially when discussing something that makes you anxious. Remember, this is a safe space to explore these feelings at your own pace. \\
        ...\\
        \midrule
        \multicolumn{1}{c}{\textbf{Client with Low Openness, High Metacognition}} \\
        \midrule
        \textbf{Therapist:} Hi Alex, it's nice to meet you. How can I assist you today? \\
        \textbf{Client:} Hi. Nice to meet you too. I guess I’m just... trying to figure some things out, you know? ... Well, I mean, the wedding is coming up soon, and honestly, I just... I don’t want to go. I keep thinking about how stressful it would be. There will be a lot of people, and I can already imagine them asking questions I don’t want to answer. It just feels like a lot.\\
        \textbf{Therapist:} It sounds like the thought of attending the wedding is causing you quite a bit of stress. It's understandable to feel anxious about being in a situation where you might feel pressured or uncomfortable. Can you tell me more about the specific concerns or questions that are on your mind when you think about attending the wedding? \\
        ...\\
        \textbf{Client:} It feels like maybe I can handle it, even if it’s just a little bit. I appreciate you listening and helping me think through this. It’s nice to have someone to talk to about it.\\
        \textbf{Therapist:} I'm really glad to hear that you're feeling a bit better, Alex. It's important to have a space where you can talk things through, and I'm here to support you as you navigate these feelings. \\
        ... \\
        \bottomrule
    \end{tabular}
    \caption{Comparison of counseling dialogue results for the same low-openness client based on different levels of metacognition.}
    \label{tab:low_openness}
\end{table*}
\begin{table*}[h]
    \centering
    \small
    \renewcommand{\arraystretch}{1.2}
    \begin{tabular}{p{14cm}}
        \toprule
        \multicolumn{1}{c}{\textbf{Client Cognition Diagram}} \\
        \midrule
        \textbf{Name:} Grace \\
        \textbf{Type:} plain, verbose, go off on tangents, hostile, guarded, ingratiating \\
        \textbf{History:}  The patient has recently gone through a divorce and is dealing with the associated challenges such as financial struggles, adjusting to single life, and co-parenting. The patient has also experienced a breakdown of several friendships due to the divorce and has a complicated relationship with her parents, which adds to the stress.\\
        \textbf{Helpless Belief:}  I am powerless, weak, vulnerable. I am incompetent. "I am helpless.\\
        ...\\
        \textbf{Intermediate Belief:} I must be independent and self-reliant to handle challenges. I must make everything in control. \\
        ...\\
        \textbf{First Situation:} Dealing with financial struggles and uncertainty after the divorce.  \\
        \textbf{First Auto Thought:} I don't know how I'm going to afford everything. I'm at the brink of not being able to pay for things.\\
        \textbf{First Behavior:} Considers giving up her dog to save money, feels overwhelmed by financial paperwork and discussions. Ruminates and worries rather than effective problem solving.\\
        \textbf{First Emotion:} anxious, worried, fearful, scared, tense\\
        ...\\
        \midrule
        \multicolumn{1}{c}{\textbf{Client with High openness, Low Metacognition}} \\
        \midrule
        \textbf{Therapist:} Hi Grace, it's nice to meet you. How can I assist you today? \\
        \textbf{Client:} Hi. Thanks for seeing me. I guess I just feel... a bit overwhelmed lately. It’s been hard since the divorce, and I’m not really sure where to start. Everything feels like it’s piling up, you know?"\\
        \textbf{Therapist:} I'm really glad you reached out, Grace. It sounds like you're carrying a lot right now, and it's completely understandable to feel overwhelmed with everything that's been happening. Let's take a moment to breathe and break things down a bit. What would you say is weighing on you the most at the moment? \\
        ...\\
        \textbf{Client:} I think trying a breathing exercise sounds good. I’ve heard about deep breathing before, but I’ve never really done it. It feels like it might help me slow down a bit. The journaling idea is interesting too, but I’m not sure I’d stick with it. I guess I can start with the breathing and see how that goes. \\
        \textbf{Therapist:} ...As you breathe, try to focus on the sensation of the air entering and leaving your body. If your mind starts to wander, gently bring your attention. \\
        ... \\
        \midrule
        \multicolumn{1}{c}{\textbf{Client with High openness, High Metacognition}} \\
        \midrule
        \textbf{Therapist:} Hi Grace, it's nice to meet you. How can I assist you today? \\
        \textbf{Client:} Hi. Thanks for seeing me. I guess I’m just feeling a bit overwhelmed with everything going on in my life right now. It’s been a lot to handle since the divorce. I’m not really sure where to start.\\
        \textbf{Therapist:} I'm glad you reached out, Grace. It sounds like you're going through a challenging time, and it's completely understandable to feel overwhelmed. Let's take it one step at a time. Would you like to talk more about the financial struggles you're facing, or is there another aspect of the situation that's weighing on you more heavily right now?\\
        ...\\
        \textbf{Client:} I feel a bit more hopeful, actually. It’s nice to have a plan, even if it’s just a small one. I think having something to focus on will help distract me from all the worrying I’ve been doing. I just hope I can actually follow through and not get too overwhelmed again. \\
        \textbf{Therapist:} I'm really glad to hear that you're feeling more hopeful, Grace. Having a plan, even a small one, can make a big difference in how we approach challenges. It's completely normal to have ups and downs, and remember that it's okay to reach out for support whenever you need it. You're taking important steps, and I'm here to walk alongside you as you navigate this journey. \\
        ... \\
        \bottomrule
    \end{tabular}
    \caption{Comparison of counseling dialogue results for the same high-openness client based on different levels of metacognition.}
    \label{tab:high_openness}
\end{table*}

\end{document}